\documentstyle[aps,prl,twocolumn,psfig,floats]{revtex} 
\def\Journal#1#2#3#4{{#1} {\bf #2}, #3 (#4)}

\def\NPB{Nucl. Phys. {\bf B}$\!$}

\def\PLB{Phys. Lett. {\bf B}$\!$}

\def\PRL{Phys. Rev. Lett.}
\def\PRD{Phys. Rev. {\bf D}$\!$}

\def\MPLA{Mod. Phys. Lett. {\bf A}$\!$}
\def\PREP{Phys. Rep.}

\def\etal{{\sl et al.}} 
\begin{document}
\wideabs{
\title{Can supersymmetric loops correct the fermion mass 
relations in SU(5)?}
\author{ J. Lorenzo Diaz-Cruz$^{1,2}$
Hitoshi Murayama$^
{1,3}$ and Aaron Pierce$^{1,3}$} 
\address{  
1. Theory Group, 
Lawrence Berkeley National Laboratory, 
Berkeley, CA~94720, USA. \\
2. Instituto de Fisica, BUAP, Puebla Pue, Mexico.\\
3. Department of Physics, 
University of California, 
Berkeley, CA~94720, USA.
}
\date{\today}
\maketitle

\begin{abstract}
  We investigate three different possibilities for improving the
  fermion mass relations that arise in grand unified theories (GUTs).
  Each scenario relies on supersymmetric loop effects alone, without
  modifying the naive Yukawa unification.  First, we consider $A$-terms
  that follow the usual proportionality condition.  In this case SUSY
  effects can improve the mass relations, but not completely.
  Interestingly, imposing Yukawa coupling unification for {\it two}
  families greatly constrains the range of parameters in the MSSM.
  Secondly, we employ a new ansatz for the tri-linear $A$-terms that
  satisfies all experimental and vacuum stability bounds, and can
  successfully modify the mass relations. Finally, we investigate the
  use of general (non-proportional) A-terms, with large off-diagonal
  entries.  In this case flavor changing neutral current (FCNC) data
  present an important constraint.  We do not pretend to present a
  complete, motivated theory of fermion masses. Rather this paper can
  be viewed as an existence proof, serving to show that Yukawa
  coupling unification can occur even within the framework of minimal
  GUTs.
\end{abstract}
}
\narrowtext



\section{Introduction} 

Unified theories that incorporate supersymmetry have met with great
success \cite{susyrev}.  Included in the attractive features of such
theories are: the possibility of ameliorating the hierarchy problem, a
natural radiative mechanism for electroweak symmetry breaking, and a
candidate for dark matter.  The most notable success of the
Supersymmetric Standard Model  (MSSM) is the successful unification of
the gauge couplings \cite{murayarev}.

In spite of all these attractive features, the minimal models appear
to be far from complete.  Most notably, the flavor sector of such
theories remains a puzzle.  One may decrease the number of mysterious
parameters in the flavor sector by placing quarks and leptons in the
same multiplets, as required in Grand Unified Theories (GUTs).  This
reduces the number of independent Yukawa couplings at the GUT scale
\cite{susygut}.  However, the predictions of such a scheme have met
with mixed success.  For instance, in the minimal $SU(5)$ implies
the tree-level relation $m_{d_i}=m_{l_i}$ at the GUT scale, which
seems to work quite well for the third generation.  However the
situation for the lighter families appears catastrophic, since we
would expect the relation $m_e/m_\mu = m_d/m_s$, but this is off by
almost one order of magnitude.  

This problem has led some authors to study non-minimal models
\cite{ellisgaill,georgijarl,halletal}, or even to question the whole
idea of grand unification \cite{antigut}. One way to solve this
problem is to invoke higher-dimensional operators \cite{ellisgaill},
an approach that essentially postpones the problem to energies beyond
$M_{GUT}$.  Another approach involves the use of non-minimal
representations for the Higgs sector.  For instance, in the context of a SU(5) GUT, by enlarging the
Higgs sector with a {\bf 45} representation, Georgi and Jalrskog
\cite{georgijarl} obtained the relations $m_d=3m_e$ and $m_s=m_\mu
/3$, which correctly reproduces the observed fermion masses when
evolved down to low energies.

In this paper we present a simpler approach which has been overlooked
thus far. Instead of adding extra matter or higher-dimensional
operators to the theory, we work within the MSSM framework.  We use a
radiative mechanism to correct the light generation mass relations,
i.e. $m_{d}=m_{e}, \, m_s=m_{\mu} $, in a manner that respects the
successful relation $m_b=m_{\tau}$. It is an exciting possibility that
one could accommodate the unification of the Yukawa couplings within
the simplest of SUSY theories, the MSSM.

The correction of fermion masses by SUSY loops has been known for a
while \cite{radfmin}.  These corrections have been applied in an
attempt to generate fully the quark and lepton masses
\cite{allradfm,nimarad}. The idea here is to use these radiative
effects in a somewhat more modest way, to fix up the GUT relations, a
task which we find to be non-trivial.  We will investigate this idea
within three contexts.  First, we investigate the case of proportional
$A$-terms.  Then, we examine two cases with non-minimal $A$-terms.  In
our second approach, we use a particular {\it ansatz} that achieves
unification for all three families. In the final approach, we
recognize that $A$-terms that are slightly more generic than the
proportional case can also help remedy the mass relations.  That is to
say, if one allows off-diagonal $A$-terms, one can remedy the Yukawa
relations.  It appears that such a scenario is still consistent with
flavor-changing neutral current (FCNC) bounds.

\nopagebreak
\section{Standard $A$-term Corrections}

We are interested in evaluating the corrections to fermion masses
coming from SUSY loops.  First, we consider the standard case where we
have $A$-terms that are proportional to the Yukawa matrices.  The
dominant corrections are the flavor conserving (FC) gluino and
Higgsino-mediated loops.  We include only these corrections in our
analysis.  Since these corrections depend on only a few MSSM
parameters it is feasible to perform a detailed numerical analysis.
Because these corrections also modify the CKM mixing matrix, we prefer
to follow a procedure consistent with all low-energy data. Our chosen
method is essentially a bottom-up approach, and consists of these
steps:
\begin{itemize}
\item Working in a basis where lepton and u-type quark mass matrices
  are diagonal, we obtain the non-diagonal d-quark mass matrices
  $(m_d)_{ij}$, at the $M_Z$-scale.  If we assume that the mass matrix
  is symmetric, we can write:
\begin{equation}
m_d= V_{CKM} \cdot m^{diag}_d \cdot V^{T}_{CKM},
\end{equation}
where $V_{CKM}$ denotes the quark mixing matrix.  The lepton mass
matrix, $m_l$, is diagonal. We can extract the quark mass matrix at
$M_Z$ utilizing the quark masses in Table I, by taking into account
the QCD renormalization effects.

\item We then modify the mass matrix of the down-type quarks by
  subtracting the contributions of the SUSY loops.  The SUSY
  corrections to the lepton mass matrix are negligible.  The flavor
  conserving (FC) corrections to the down quarks includes the gluino
  and Higgsino loops, namely $(\delta m_d)^{FC}= (\delta m_d)^{\tilde
    g}+(\delta m_d)^{\tilde H}$.
  
\item The Yukawa couplings are simply related to the mass matrix
  through the vacuum expectation value of the down-type Higgs.
  Namely,
\begin{equation}
Y^{d}=\frac{m_{d}}{v \sin \beta},
\end{equation}
where $v$=174 GeV.  We convert the Yukawa couplings from the
$\overline{MS}$ scheme to the $\overline{DR}$ scheme, using the
dictionary of Martin and Vaughn \cite{MStoDR}. This is necessary to be
compatible with the renormalization group equations used in the next
step.

\item Once we have the Yukawa couplings of both quarks and leptons at
  $M_{Z}$ in hand, we evolve the Yukawa matrices up to the GUT scale
  ($M_{GUT}$), using the renormalization group equations of the MSSM
  \cite{MSSMRGE}. This yields the Yukawa matrices at the GUT scale
  $Y^d(M_{GUT})$ and $Y^l(M_{GUT})$.
  
\item The next step is to diagonalize the Yukawa matrix,
  $Y^d(M_{GUT})$, and compare its eigenvalues with the lepton Yukawa
  couplings at the GUT scale, $Y^l(M_{GUT})$, to see whether
  unification has occurred. To quantify this unification we define
  $\epsilon_{i}= |(Y^d_{ii}-Y^l_{ii})/Y^l_{ii}|$.
  
\item We repeat the above steps, scanning the parameter space of the
  MSSM, to search for those regions where $\epsilon_{i}$ is close to
  zero.
\end{itemize}

\begin{table}
\label{tab:inp}
\begin{tabular} {l|l} 
Input & Value ($\overline{MS}$ scheme) \\ \hline
$\alpha_{s}(M_{Z})$ & 0.119 \\ \hline
$\sin \theta_{W}$ & 0.23124   \\ \hline
$\alpha_{EM}^{-1}(M_{Z})$ & 127.943    \\ \hline
$m_t(m_t)$ & 165 GeV \\ \hline
$m_b(m_b)$ & 4.25 GeV\\  \hline
$m_c(m_c)$ & 1.15 GeV\\ \hline
$m_s(\mbox{ 2 GeV})$ & 115 MeV \\ \hline
$m_d(\mbox{ 2 GeV})$ & 6 MeV\\ \hline
$m_u(\mbox{ 2 GeV})$ & 3 MeV \\ \hline
$m_{\tau}(m_{\tau})$ & 1.777 GeV\\ \hline
$m_{\mu}(m_{\mu})$  & 105 MeV\\ \hline
$m_{e}(m_{e})$ &  0.511 MeV \\ \hline
\end{tabular}
\caption{Inputs for the numerical analysis of section II.}
\end{table}

Now that we have outlined the steps in the analysis, we note the
explicit corrections to the Yukawa couplings.  The gluino loop,
$(\delta m_d)^{\tilde g}$, contributes to all the diagonal elements
$m_{dd,ss,bb}$.  We work in the limit when $\mu\tan\beta$ is much
larger than the contribution from the tri-linear scalar coupling.  In
this limit, the gluino loop contribution may be expressed as
\cite{hallratsa}:
\begin{equation}
 (\delta m_d)^{\tilde g}_{ii} = \frac{2\alpha_s}{3\pi}
 \frac{\mu M_{\tilde g}}{M^2_X} I(x) \tan\beta m_{d_i}. 
\end{equation}
The Higgsino-mediated loop $(\delta m_d)^{\tilde H}$, is only relevant
for the third family due to the suppression from small Yukawa
couplings.  Its contribution is given by:
\begin{equation}
 (\delta m_d)^{\tilde H}_{ij} = \frac{Y_t}{4\pi}
 \frac{\mu A_t}{M^2_X} I(x) \tan\beta  \delta_{i3}\delta_{j3}.
\end{equation}
As usual, $M_{\tilde g}$ denotes the gluino mass, $\mu$ is the
parameter for the Higgs term in the superpotential, and $A_{t}$
denotes the tri-linear scalar term for top. $M_X$ denotes the largest
mass appearing in the loop, and $I(x)$ denotes the loop integral,
given by
\begin{equation}
\label{loopintegral}
I(x)= \frac{-1+x-x \log(x)}{(1-x)},
\end{equation}
where $x=m^2_{\tilde g}/m^2_{\tilde q}$ for the gluino loop, and
$x=\mu^2/m^2_{\tilde q}$ for the higgsino loop.

The key point in this analysis is to recognize that the higgsino and
gluino pieces come with an opposite sign.  As a result these two
contributions may nearly cancel each other for the third generation,
thereby preserving the $m_{b}=m_{\tau}$ relation to some extent.
However, since the higgsino piece only contributes to the third
generation, the gluino piece remains uncancelled for the other two
families and can be utilized to fix up a mass relation.  Of course, it
can only be used for one of the two families.  It will exacerbate the
problem for the remaining family.

The results for the parameter $\epsilon_{i}$ for the 3 generations are
shown in FIG. 1-3. These graphs correspond to $m_{\tilde g}=m_{\tilde
  q}=500$ GeV, and $\mu=-600$ GeV.  Notice that there is a limited
region in the $(A_{t},\tan \beta)$ plane where unification (say
$|\epsilon| < .2 \%$) occurs for both the the first and third
generation.  In fact, if we assume that this approach is part of the
solution to remedying the mass relations, we can view the results of
this analysis as making predictions for tan $\beta$ and $A_{t}$.  We
simply mention here that the other sign of $\mu$ allows one to achieve
unification for the second and third generation.  Though this appears
to be more difficult, and requires large values of $\tan \beta$ where
the peturbativity of the b-quark Yukawa coupling is an issue.

One could view this scenario with some wariness, as it remedies the
mass relation for one generation at the expense of exacerbating it for
the other generation.  Nevertheless, this sort of scenario allows one
to get the correct fermion spectrum by, for instance, introducing just one
higher-dimensional operator involving the remaining family.  In some
sense, we have achieved ``flavor from no flavor''.

\begin{figure}[h!]
\psfig{file=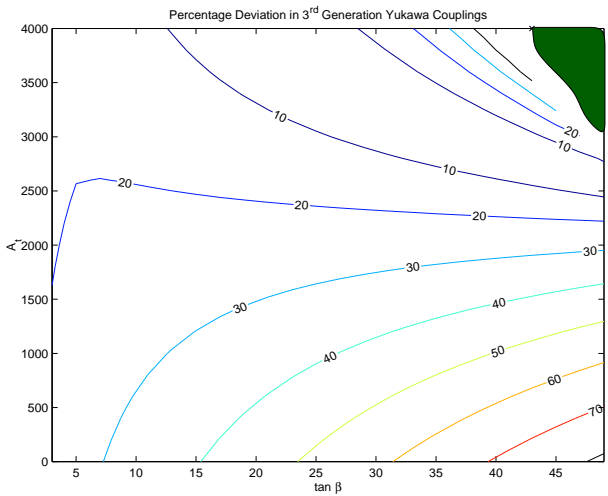,width=3.3in}
\caption{The percentage deviation in Yukawa couplings, $100 \times
  \epsilon_{i} \equiv 100 \times
  |(Y^d_{ii}-Y^l_{ii})/Y^l_{ii}|$, for the third generation. The
  dark shaded region represents the area where the b quark Yukawa
  coupling is non-perturbative, and the analysis breaks down.  The
  figure corresponds to $\tilde{m_{g}}=500$ GeV and $\mu=-600$ GeV.} 
\end{figure}

\begin{figure}[h!]
  \psfig{file=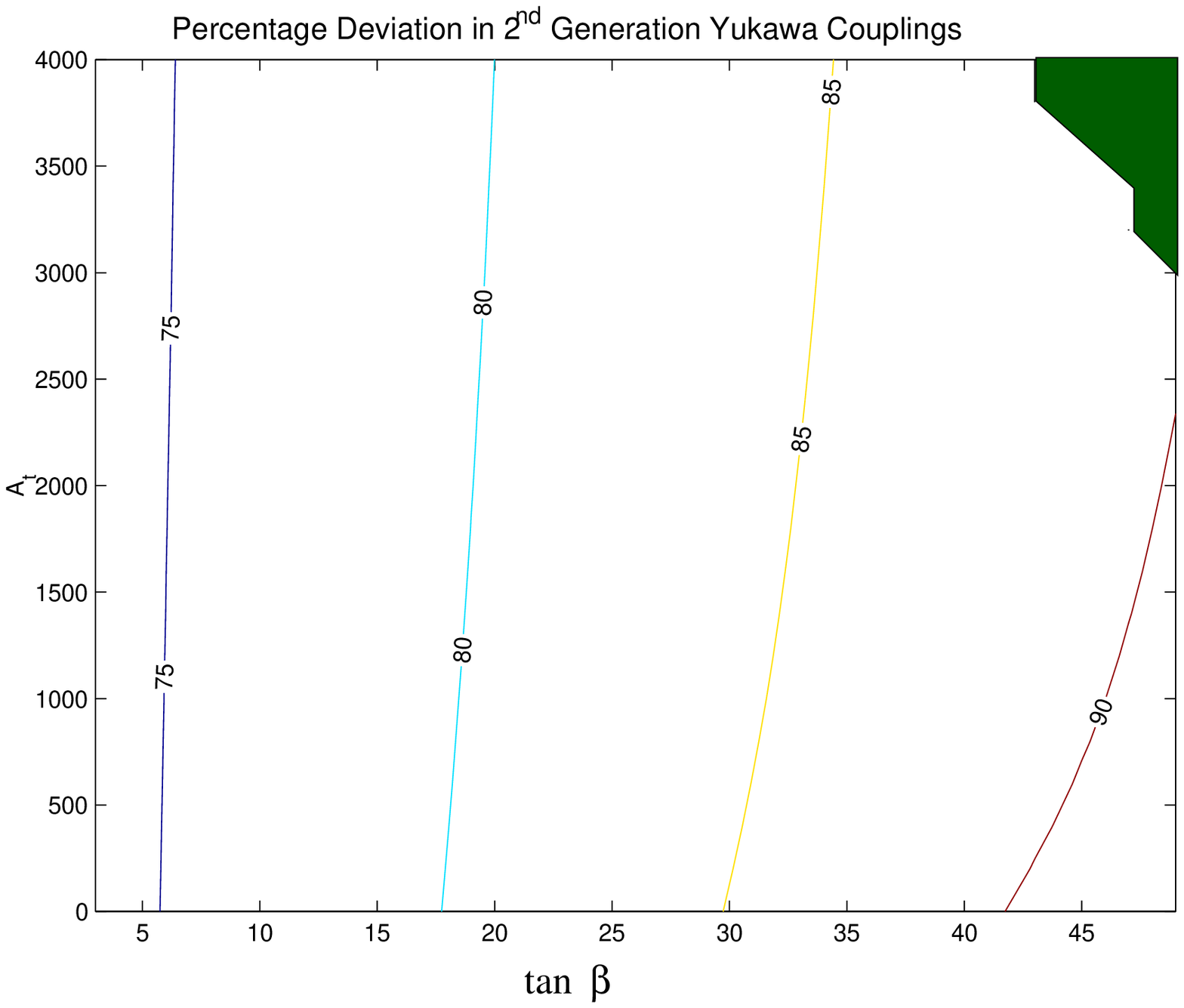,width=3.3in}
\caption{The percentage deviation in Yukawa couplings, $100 \times
  \epsilon_{i} \equiv 100 \times |(Y^d_{ii}-Y^l_{ii})/Y^l_{ii}|$, for
  the second generation. The dark shaded region represents the area
  where the b quark Yukawa coupling is non-perturbative, and the
  analysis breaks down.  The figure corresponds to $\tilde{m_{g}}=500$
  GeV and $\mu=-600$ GeV.} 
\end{figure}

\begin{figure}[h!]
\psfig{file=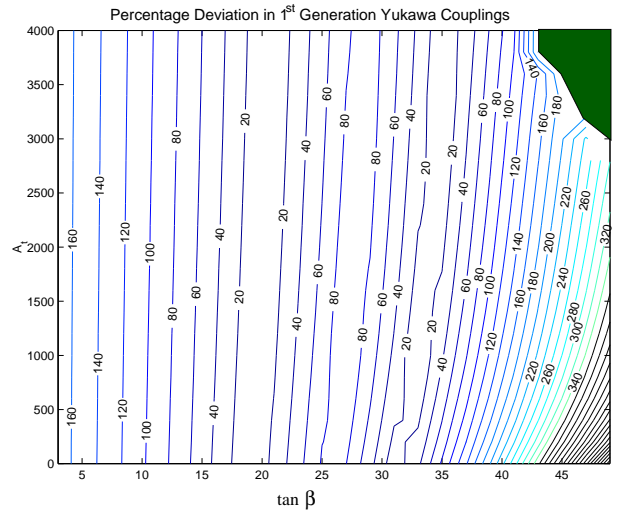,width=3.3in}
\caption{The percentage deviation in Yukawa couplings, $100 \times
  \epsilon_{i} \equiv 100 \times |(Y^d_{ii}-Y^l_{ii})/Y^l_{ii}|$, for
  the first generation. The dark shaded region represents the area
  where the b quark Yukawa coupling is non-perturbative, and the
  analysis breaks down. The figure corresponds to $\tilde{m_{g}}=500$
  GeV and $\mu=-600$ GeV.}   
\end{figure}

\section{A non-minimal ansatz for the $A$-terms.}

Since our initial efforts with proportional $A$-terms did not succeed
in achieving complete unification, we are forced to resort to a more
general form for the $A$-terms, thereby opening the door for a larger
number of free-parameters. This makes it more difficult to perform a
detailed numerical analysis.  Instead, we shall be content to present
an ``existence proof" of the viability of these mechanisms.  In the
following, we shall employ a top-down approach, by evolving the Yukawa
couplings from the GUT-scale to the weak-scale, and then including the
threshold corrections.

We shall begin by considering a particular {\it ansatz} for the
tri-linear terms, aimed at correcting the Yukawa matrices for the
light generations. In this section, we say nothing about $b$-$\tau$
unification, since that relation is already close, and presumably can
be fixed up by methods similar to those used in the previous section.
So, let us start by writing the down-type fermion mass matrix of first
and second generations as:
\begin{equation}
\label{massmat}
M^0=  \left(
\begin{array}{ll}
 0             &   \lambda m \\
 \lambda m     &   m 
\end{array}
\right) 
\end{equation}
Thus, at GUT scale one has $m^0_s=m^0_\mu=m$, and
$\lambda^2=m^0_e/m^0_\mu=m^0_d/m^0_s$.  The superscript 0 will
indicate fermion masses evaluated at the GUT scale.  Now, we shall
include the correction coming from the gaugino-sfermion loops, using
the an {\it ansatz} for the $A$-terms that results in the following
contribution to the squark mass matrices:
\begin{equation}
\label{ansatzmatrix}
\tilde{m}^{2}_{LR}  \supset  \left(
\begin{array}{ll}
 0    &   0 \\
 0     &  X 
\end{array}
\right) v \cos \beta + \left(
\begin{array}{ll}
 0    &   \lambda \\
 \lambda     &  1 
\end{array}
\right) A Y_{s} v \cos \beta
\end{equation}
Here $Y_{s}$ is the Yukawa coupling for the strange quark.  We have separated out the first piece to emphasize the largeness
of the tri-linear term, $X$, that is necessary as we will see below.
The grand unification forces the term $X$ to be common to both charged
leptons and down-type quarks.  We note that in the mass-eigenstate
basis for the fermions, the first matrix on the right-hand side
becomes off-diagonal, which can lead to FCNC problems.  We will
address this issue shortly.  Despite our concerns about FCNCs, and the
fact that we do not yet have a microscopic theory to motivate our {\it
  ansatz}, we will boldly forge ahead.  We now show that this form for
the $A$-terms proves to be phenomenologically very interesting in
repairing the mass relations.

The dominant correction to the quark masses arises from the
gluino-mediated loop.  This is larger than the bino-loop contribution
due to the difference in the gauge couplings. Therefore, for questions
of unification, we can again neglect the corrections to lepton masses.
Moreover, because of the form chosen for the tri-linear coupling
terms, Eqn.\ (\ref{ansatzmatrix}), the gluino-loop contributes mostly
to the (2,2) entry of the quark mass matrix.  For simplicity, we
neglect all corrections but this one.  After this correction, the
lepton mass matrix approximately retains the from of Eqn.\ 
(\ref{massmat}), while the quark mass matrix becomes:

\begin{equation}
\label{masscorr}
M_{d} =  \left(
\begin{array}{ll}
 0    &   \lambda m \\
 \lambda m     &   m + \delta m_{s} 
\end{array}
\right) 
\end{equation}

The basic idea will be to utilize the gluino loop correction, $\delta
m_{s}$, to cancel a substantial portion of the muon mass, $m$, derived
from the Yukawa coupling.
 

To determine the precise amount of cancellation necessary, one must
look at the experimental data at low energy.  In particular, given the
mass matrix of the form in Eqn.\ (\ref{masscorr}), one generates a
Cabibbo angle, $\lambda_c$ given by the relationship:
\begin{equation}
\label{cabbibo}
\lambda^2_c=\frac{m_d}{m_s}.
\end{equation}
We note that the masses in this equation are the physical masses,
which yields $\lambda_c \approx 0.22$.  This value is consistent with
other determinations of $\lambda_{c}$, a well-known fact.  Now,
utilizing the diagonalized form of Eqn.\ (\ref{masscorr}) for the
corrected mass matrix, we can write:
\begin{equation}
\label{fermionfixer}
\lambda_{c} \approx \frac{ \lambda  m^{Z}_{s}}{(m^{Z}_{s}+\delta m_{s})/(1+\lambda_{c}^2)}.
\end{equation}
We also note that the ratio $\lambda^2=m^0_e/m^0_\mu$ is independent of scale (to one loop), so we can fix the value of  $\lambda$ at the weak-scale using the physical lepton masses, namely $\lambda \sim 0.07$.  This gives us the relation:
\begin{equation}
\label{fermionrequirement}
1 + \frac{\delta m_{s}}{m_{s}^{Z}} \approx \frac{1}{3}. 
\end{equation}
Also, we note that, after RGE running, we have the approximate relation, $m^{Z}_{s} \approx 2.1 m_{\mu}$. 

Finally, we wish to determine the constraint on the SUSY parameters.  To do this,  we need to relate the SUSY parameters to $\delta m_{s}$. The correction due to the gluino loop can be written as \cite{hallratsa}:
\begin{equation}
\label{Acorrect}
 (\delta m_s) = \frac{2\alpha_s}{3\pi}
 \frac{m^2_{LR} M_{\tilde g}}{M_{X}^2} I(x) , 
\end{equation}
where $m^2_{LR}$ is given in Eqn.\ (\ref{ansatzmatrix}). We assume
that $m^2_{LR}$ is completely dominated by the piece proportional to
$X$.  Here, $M_X$ denotes the largest mass appearing in the loop, and
$I(x)$ denotes the loop integral, given in Eqn.\ (\ref{loopintegral}),
which becomes close to 1 in magnitude at small $x$.  Indeed, we will
be considering small $x$, as we will see $m^{2}_{\tilde{l}} \approx
m^{2}_{\tilde{q}} \gg M_{\tilde{g}}^{2}$, is necessitated by the FCNC
data coupled with concerns about naturalness.  In the following, we
denote the mass of all heavy scalar particles by $\tilde{m}$.

Therefore, combining Eqs.\ (\ref{ansatzmatrix}),
(\ref{fermionrequirement}), and (\ref{Acorrect}), we find that the
supersymmetric parameters must satisfy the following constraint if we
are remedy the mass relations:
\begin{equation}
\label{massrequirement}
\frac{M_{\tilde{g}} X \cos \beta}{\tilde{m}^2}  \approx 0.035.  
\end{equation}

As we intimated earlier, it is a non-trivial task to reconcile this
requirement with FCNC data.  In particular, a strong constraint comes
from the $\mu \rightarrow e \gamma$ limit \cite{mega}.  We find that
once a gluino mass is chosen, Eq.\ (\ref{massrequirement}), along with
the limit from $\mu \rightarrow e \gamma$, sets the mass scale for the
sleptons.  For the gluino, we choose a mass of 600 GeV, and utilize
the relation $M_{\tilde{g}} \approx 7 M_{1}$ that arises from the
unification of gaugino masses.  A rotation of the matrix of Eq.\ 
(\ref{ansatzmatrix}) to the diagonal fermion mass basis generates a
$\delta_{12}$ in the lepton sector given by

\begin{equation}
\label{modelinsertion}
(\delta_{12}^{l})_{LR} \sim \frac{X \cos \beta}{\tilde{m}^2} \lambda v
= \left(\frac{M_{\tilde{g}} X \cos \beta }{\tilde{m}^2}\right)  
\frac{\lambda v}{M_{\tilde{g}}} = 7.1 \times 10^{-4}.
\end{equation}
Here, we have inserted the value for $\lambda$ and used Eqn.\ 
(\ref{massrequirement}).  Using this value for
$(\delta_{12}^{l})_{LR}$ and the constraint on the $\mu \rightarrow e
\gamma$ branching ratio, one can find $\tilde{m} \stackrel{>}{\sim}$
4.4 TeV.

It then remains to check whether a scenario with 4.4 TeV sleptons and
squarks satisfies the remaining FCNC data.  Note that the expression
for $(\delta^{d}_{12})_{LR}$ is identical to Eq.\ 
(\ref{modelinsertion}), with the replacement $\lambda \rightarrow
\lambda_{c}$.  Using the analysis of \cite{deltamk}, which includes
NLO QCD effects, we have checked that such a scenario satisfies the
bounds from $\Delta M_{K}$ by about a factor of 5.  However,
constraints from $\epsilon$ \cite{deltamk}, and $\epsilon^\prime /
\epsilon$ \cite{masieroetal,KTeV,NA48}, are not satisfied unless the
phase of $(\delta^{d}_{12})_{LR}$ is less than about 0.2.  We have
also checked, adapting the analysis of \cite{moroi}, that this
scenario is safe from measurements of the anomalous magnetic moment of
the muon\cite{mugminus2}.
  
Incidentally, considering the large value of the tri-linear coupling
required in this scenario, one might be concerned about charge and
color breaking minima of the scalar potential.  For the values of
$M_{\tilde{g}}$ and $\tilde{m}$ described above, with $\tan \beta
\sim 2$, we have $X \simeq 2500$.  Although it is possible that such a
scenario might not have the the correct electroweak symmetry breaking
(EWSB) vacuum as an absolute minimum, $X \simeq \tilde{m}$ does allow
the EWSB vacuum satisfying the less stringent criteria of
meta-stability \cite{borzumetal}.  That is to say, we succeed in
producing a EWSB vacuum possessing a lifetime greater than the age of
the universe.

Finally, we wish to comment on proton decay in this scenario.  One might expect the proton decay rate to be significantly affected on two counts.  First of all, the first and second generation scalar masses are pushed to several TeV.  However, increasing the masses for the first and second generation superpartners is not enough to decrease the proton decay rate predicted by SU(5) below experimental bounds, a point implicit in \cite{GotoNihei}, and later emphasized in reference \cite{MurayamaPierce}.  This is due to the presence of an important contribution arising from the exchange of third generation superpartners that remains even if the first two generations decouple.  Secondly, one might be concerned that the dimension 5 operators, which are proportional to Yukawa couplings, might be significantly modified.  However, since the dimension 5 operators are proportional to the $(2,1)$ element of the Yukawa matrix, our large correction to the $(2,2)$ element is not significant.  Finally, one might be interested in entirely new contributions to proton decay arising from the new large A-term that might somehow effect a cancellation.  However, we have checked that such contributions are at most sub-dominant.  So, the proton decay rate in this scenario is comparable to the usual supersymmetric SU(5) GUTs, which admittedly have difficulty accommodating the experimental bounds.

Finally, it should be noted that the unification of fermion masses with this {\it ansatz} requires both
scalars of a few TeV, while gauginos may remain relatively light.
Many possibilities exist for such scenarios \cite{decoupling}.
However, none of these scenarios generates a large enough tri-linear
coupling, $X$.  Until a micro-physical motivation can be found for
this case, the above discussion must be simply regarded as an
existence proof.

\section{Corrections From Flavor-Changing Gluino Interactions}

Finally, let us consider a scenario in which we allow effects from
flavor violating (FV) gluino-quark-squark interactions to enter into
our analysis.  Such contributions are known to induce potentially
dangerous FCNC effects.  Therefore, we must be careful to satisfy such
bounds.

In this case, the mass-insertion method proves somewhat complicated,
as several insertions are necessary.  Therefore, we find it convenient
to write the vertex $\tilde{g}$-$\tilde{d}_{L,R}$-$d_{L,R}$, in terms
of $3 \times 3$ mixing matrices,$W_{L,R}=V^d_{L,R}
\tilde{V}^{d\dagger}_{L,R}$,where $V^d_{L,R}
(\tilde{V}^{d\dagger}_{L,R})$ denote the rotation matrices required to
diagonalize the fermion (sfermion LL,RR) mass matrices.  A discussion
of the bounds on the $W_{L,R}$ matrices is presented in
\cite{nimarad}.

In general, the corrections to fermion masses will include corrections
from both gluino and Higgsino mediated loops. In the large $\tan\beta$
limit, it is reasonable to include the Higgsino loop only for the
heavier fermions (third generation), whereas the gluino flavor
conserving (FC) loop contributes to all diagonal elements of the
fermion mass matrices.  The size of the gluino loop is proportional to
the fermion mass itself.  Thus, we can write the correction to quark
masses as: $\delta m_d= (\delta m_d)^{FC} + (\delta m_d)^{FV}$.  For
the quarks, $(\delta m_d)^{FC}$ is given in the previous section in
Eqn.\ (\ref{Acorrect}), the FV gluino contribution is given by
\cite{nimarad},
\begin{equation}
 (\delta m_d)^{FV}_{ii} = 0.7 m_b \frac{\mu}{M_{\tilde g}}
 \frac{\tan\beta}{60} \frac{x^{1/2} \tilde{H}}{0.5}  T^d_{3i}
\end{equation}
where $T^d_{3i}= W^d_{L,3i} W^{d\dagger}_{R,3i}$, $\tilde H$ denotes a
combination of loop integrals, analogous to Eqn.\ 
(\ref{loopintegral}), and in general $\tilde H \simeq 0.5$.  Note that
this equation is proportional to $m_b$, which will allow these
corrections to reach substantial values, a prerequisite for correcting
the GUT relations.

In order to obtain the required values for the fermion masses, it
seems convenient to fix the values of the FC and FV contributions, in
such a way that their contribution to $m_b$ cancels.  Again, we
neglect the contribution to the leptons.
In this case, the ratios $m_{d_i}/m_{l_i}$ can be written as:
\begin{equation}
\frac{m_{d_i}}{m_{l_i}}= \left(1+\frac{\delta m_{d_i}}{m_{d_i}} \right)
\frac{m^0_{d_i}}{m^0_{l_i}}. 
\end{equation}
Here, the 0 superscript again denotes evaluation at the GUT-scale.  In
the above expression,$\delta m_{d_i}$ includes contributions from both
FC and FV pieces.  In order to generate the correct mass ratios,
$\delta m_{d_i}/m_{d_i}$ must in turn satisfy:
\begin{equation}
\left(\frac{\delta m_d}{m_d}\right)^{FC}+
\left(\frac{ \delta m_d}{m_b}\right)^{FV}\frac{m_b}{m_d}=2
\end{equation}
\begin{equation}
\left(\frac{ \delta m_s}{m_s}\right)^{FC}+
\left(\frac{ \delta m_s}{m_b}\right)^{FV}\frac{m_b}{m_s}=-\frac{2}{3}
\end{equation}
After substituting the expressions for the factors resulting from Eqn.
(8), one finds that the correct mass ratios require: $T^d_{31}
T^d_{32} \simeq 10^{-5}$ (for $\tan\beta=45$).  This solution is in
agreement with current bounds on FCNC phenomenology, which imply:
$T^d_{31} T^d_{32} < 2 \times 10^{-5}$ \cite{nimarad}.\footnote{In
  fact, there are no direct bounds on each factor $T^d_{31,32}$;
  however, by combining present bounds on several FCNC processes
  involving the $W^d_{L,R}$ elements, one can arrive at the quoted
  bound} Thus, the quark-lepton mass relations for all three families
can be remedied, using a large effect from both FC and FV corrections.

There could be some concern about whether the Yukawa coupling of
the b quark would remain perturbative in such a large $\tan \beta$ scenario, or whether this solution could survive after more stringent bounds on FCNC are applied. In the face of such objections, it should be noted that including the FV corrections to the lepton masses
(which will be proportional to the tau mass and could conceivably be non-negligible), or by scanning the relevant parameters, one should be able to retain this  (or a similar) solution.

\section{Conclusions}

The main result of this paper is that the ``wrong fermion mass relations'' of the minimal SUSY GUTs can be greatly ameliorated by employing radiative corrections due to the superpartners of the SM particles. 

Employing the conventional $A$-terms, we can obtain correct unification relations for the Yukawa couplings for two families, depending on the sign of the $\mu$ parameter. We call this scenario ``flavor from no flavor''.  This scenario effectively limits the parameter space of the MSSM where two families correctly unify.  However, to unify the remaining family a higher dimensional operator (or a more complicated higgs sector) is required. 


We have also presented a non-minimal ansatz for the $A$-terms, that fixes the relations $m_d=m_e,\, m_s=m_\mu$, while retaining the successful  relation $m_b=m_\tau$. This ansatz is barely consistent with current FCNC bounds.

As a final approach to modify the mass relations, we considered Flavor violating (FV)  gluino-quark-squark interactions.  The values of the A-matrix required barely satisfy all phenomenological and stability bounds of the MSSM.

Although this paper does not present a theory of fermion masses,
it does presents three cases that can be considered as ``counter-examples'' to the usual claim that the SU(5) GUT model is excluded because of the wrong fermion mass relations. We stress our result:  fermion mass relations can work for all three families even in the minimal SU(5) GUT.

\section{Acknowledgments} 
The work of J.L. D.-C. thanks the support of CONACYT
and SNI (M\'exico), as well as the kind hospitality of LBL.  The work of AP and HM was supported in part by the Director, Office of Science, Office of High Energy and Nuclear Physics, Division of High Energy Physics of the U.S. Department of Energy under  Contract DE-AC03-76SF00098 and in part by the National Science Foundation under grant PHY-95-14797.  AP is also supported by a National Science Foundation Graduate Fellowship.

\end{document}